\documentclass[article]{mesa-NS}
\usepackage{times,mathptm}
\usepackage{amsmath,amsfonts,amssymb}
\usepackage{graphicx,graphics,epsfig}
\usepackage{fancyhdr,fancybox}
\usepackage{verbatim}
\usepackage{color}

\usepackage[english]{babel}
\usepackage[utf8x]{inputenc}
\usepackage[T1]{fontenc}

\jvol{xx}
\jnum{xx}
\jyear{20xx}

\begin{document}
\label{pageinit}

\date{}


\title{The dangerous path towards your own cryptography method}

\author{Warley M.S. Alves$^1$, Thiago L. Prado$^1$, Antonio M. Batista$^2$ and Fabiano A.S. Ferrari $^{1,\star}$}

\maketitle

\noindent $^1$ Institute of Engineering, Science and Technology, Federal University of the Jequitinhonha and Mucuri's Valleys, Janaúba, Brazil.\\
\noindent $^2$ Department of Mathematics and Statistics, State University of Ponta Grossa, Ponta Grossa, Brazil.\\

\noindent $^\star$ \emph{Corresponding Author}. fabianosferrari@gmail.com

\markboth{WMS Alves, TL Prado, A Batista and FAS Ferrari}{The dangerous path towards your own cryptography method}


\footnotetext[2010]{\textit{{\bf Mathematics Subject Classification}}  }

\begin{abstract}
Would you like to have your own cryptography method? Experts say you should not do it. If you think you can develop a better cryptography method anyway. We present a brief discussion about some well known cryptography methods and how our model fails against the traditional attacks. We do not want to discourage anybody, we just want to show that, despite of the importance of developing better cryptography models, it is a very hard task.
\end{abstract}

\section{Introduction}

Everyday we receive spams and fake news, besides, everybody is subjected to hacker attacks. The Internet and smartphone's development have lead the humanity to a new level of interconnectivity. This technological advance comes with a price, nobody is completely safe when buying products and services online, or exchanging documents or personal data through communication apps. To overcome the security issues, a constant evolution of the algorithms of cryptography is required. 

Cryptography is the study of mathematical problems to solve two types of security problems: privacy and authentication \cite{Diffie1976}. One of the main concerns of cryptography is to make the message unreadable and unalterable for eavesdropper and saboteurs \cite{Diffie1976}. The cryptography consists of using an algorithm and a secret key  to encrypt a plaintext (message) into a cyphertext and recovering the plaintext using another secret key. The cryptography algorithms can be divided into two categories: symmetric key and asymmetric key \cite{Kumar2011}. In the symmetric key algorithms, the key to encrypt the plaintext into a cyphertext is the same the decrypt the cyphertext back to the plaintext. In the asymmetric key algorithms the keys to encrypt and decrypt the message are different.  

Despite of structural differences, both symmetric and asymmetric key algorithms are widely used nowadays \cite{Patil2016}. One example of symmetric key algorithm is the Advanced Encryption Standard (AES) algorithm. It was developed in 1998 and since 2001 was established by the U.S. National Institute of Standards and Technology as standard algorithm for encryption of electronic data \cite{Schwartz2000}. One of the first asymmetric key algorithms is the Rivest–Shamir–Adleman (RSA) algorithm, it was developed in 1978 and is often used for secure data transmission \cite{Patil2016}. Details about the RSA algorithm are going to be discussed in Section 2.2. 

Most of the standard cryptography algorithms are based on prime numbers factorization and modular operations. A different approach is the use of the properties of chaotic maps to encrypt secret messages. The first chaos based cryptography model was proposed in 1989 by Robert Matthews \cite{Matthews1989}. In 1991, the chaotic properties of the tent map was explored to generate a cryptography algorithm for the first time \cite{Habutsu1991}. As soon as the chaos based cryptography models were proposed, the vulnerabilities of this type of system started to be reported \cite{Biham1991}. Several chaos based cryptography model have been proposed since Matthews's algorithm \cite{Chua1997,Kanso2009,Grzybowski2009,Pizolato2011}. Proposed in 1998, the Baptista's chaos based algorithm is one of the most popular among them \cite{Baptista98}. The Baptista's algorithm is based on the ergodicity properties of the logistic map and is going to be discussed in more details in Section 2.3. 

In 2008, Hung and Hu presented for the first time the concept of exchange information using direction of coupling between chaotic maps \cite{Hung2008}. The Hung and Hu model used a quantity called Transfer Entropy to detect the direction of coupling. In this paper, we present a generalization of this concept where any quantity able to detect the direction of coupling between chaotic maps can be used. 

This paper is organized as follows: in Section 2 we discuss three very popular cryptography algorithms, in Section 3 we introduce our cryptography model, in Section 4 we demonstrate how different quantities can be used to detect presence or absence of coupling between chaotic maps, in Section 5 we present an example of how our model works, in Section 6 we discussed the robustness and security issues of our model and in Section 7 we provide our final considerations.

\section{Some known cryptography algorithms}

To make the reader familiar with the cryptography algorithms, we present three famous algorithms models. The One Time Pad (OTP) is one of the simplest and famous cryptography algorithms, the RSA algorithm is widely used for Internet purposes and the Baptista's chaos based algorithm is the most popular chaos based algorithm for cryptography. 

\subsection{One Time Pad (OTP)}

One of the most famous cryptography schemes is the One Time Pad (OTP). It was first described by Frank Miller in 1882 \cite{Bellovin11} and the patent of this method was issued by Gilbert S. Vernam in 1919 \cite{Vernam19}. It consists of a modular arithmetics that changes a message into a cyphertext. To generate a cyphertext is necessary a message $m$ of length $L$ and a secret key $k$ of the same length. Let's suppose that Alice wants to send a message to her friend Bob, she decides to send a simple message like "HELLO" that is $5$ letters long. She chooses "TODAY" as the secret key. To generate the cyphertext $c$ is necessary to apply an XOR operation (symbol $\oplus$) between the message and the secret key:
\begin{eqnarray}
c=(m \oplus k) \mbox{ mod N}, 
\end{eqnarray}
where $N$ is the size of the table of characters, 'mod' is the modular operation that means that once the sum is bigger than the size of the table then the list is reset to its beginning. Let's considered that Alice uses the English alphabet that contains 26 letters. The alphabet is presented in Table \ref{tb1}.

\begin{table}
\caption{Alphabet as a numerical sequence.} 
\begin{center}
\begin{tabular}{|c|c|c|c|c|c|c|c|c|c|c|c|c|c|c|c|c|c|c|c|c|c|c|c|c|c|}
\hline
A & B & C & D & E & F & G & H & I & J & K & L & M & N & O & P & Q & R & S & T & U & V & W & X & Y & Z \\ \hline 
1 & 2 & 3 & 4 & 5 & 6 & 7 & 8 & 9 & 10 & 11 & 12 & 13 & 14 & 15 & 16 & 17 & 18 & 19 & 20 & 21 & 22 & 23 & 24 & 25 & 26 \\  \hline
\end{tabular}
\end{center}
\label{tb1}
\end{table}

In this case the modular operation gives us the cyphertext $c$ 
\begin{eqnarray}
c&=&(\mbox{HELLO }\oplus\mbox{TODAY }) \mbox{ mod 26} ,\nonumber \\
c&=&\mbox{BTPMN }.
\end{eqnarray}
The only way Bob can decrypt the cyphertext $c$ is using the secret key $k$. In general, the cyphertext can be recovered applying the XOR operation between the cyphertext and the secret key,
\begin{eqnarray}
m=(c \oplus k) \mbox{ mod N}.
\end{eqnarray}
Consequently, 
\begin{eqnarray}
m&=&(\mbox{BTPMN }\oplus\mbox{TODAY }) \mbox{ mod 26} ,\nonumber \\
m&=&\mbox{HELLO }.
\end{eqnarray}
The OTP is an example of symmetric key algorithm.

The biggest advantage of the OTP is the fact that it is unbreakable if the secret key is used only one time. On the other hand, for each new message a new secret key needs to be generated. Another disadvantage is that the secret key needs to be as long as the message.

\subsection{RSA algorithm}

Different from the One Time Pad, the RSA method is asymmetric, this means that the key used to cypher the plaintext is different from the key used to decipher the cyphertext. The algorithm was published in 1978 by Ron {\bf R}ivest, Adi {\bf S}hamir, and Leonard {\bf A}dleman \cite{Rivest78}. The encryption-decryption scheme is based on trap-door one-way functions, these functions have the property that they are easy to compute one way but very difficult to compute in the other direction \cite{Rivest78}. 

This is an example of asymmetric key algorithm. The message $m$ is encrypted by a public key $k(e,n)$, where $e$ and $n$ are a pair of positive integers. The encryption/decryption RSA algorithm can be implemented as follow:

\begin{enumerate}
\item The message $m$ needs to be break in blocks between $0$ and $1-n$ long. 
\item The encryption/decryption will be similar to the One Time Pad
\begin{eqnarray}
c=m^e \mbox{ mod }n,\\
m=c^d \mbox{ mod }n,
\end{eqnarray}
where $d$ defines the decryption key $k(d,n)$, $d \neq e$. 
\item To define $n$ is necessary to choose two prime numbers $p$ and $q$, such as
\begin{eqnarray}
n=pq.
\end{eqnarray}
\item To define $d$ is necessary to find a random integer number that is prime of $(p-1)$ and $(q-1)$, i.e., $d$ is a parameter that satisfies the the greatest common divisor $gcd$
\begin{eqnarray}
gcd(d,(p-1),(q-1))=1.
\end{eqnarray}
\item The parameter $e$ is the multiplicative inverse of $d$, i.e.,
\begin{eqnarray}
ed \equiv 1 \mbox{ mod } (p-1)(q-1).
\end{eqnarray}
\end{enumerate}

As an example, if you choose $p=47$ and $q=59$, then $n=2773$, one $d$ value that satisfies the condition 4 is $d=157$, the parameter $e$ that satisfies condition 5 is $17$. With this set of parameters is possible to encrypt and decrypt the message using condition 2 in a safe manner. 

\subsection{Baptista's Chaos Based Algorithm}

In 1998, Murilo S. Baptista proposed a new approach for cryptography using chaotic maps \cite{Baptista98}. One map that can be used in his approach is the logistic map, defined as
\begin{eqnarray}
x_{n+1}=ax_n(1-x_n),
\end{eqnarray}
where $a$ is the bifurcation parameter and $x_n \in [0,1]$. The system is secure when the parameter $a$ is defined in the chaotic regime. 

In this method, the first step is to define how the alphabet will be coded. It can be generated dividing the size of the attractor in bins of equal size, according to the number of characters of the alphabet. Using the English alphabet, for instance, each bin will have the size $\epsilon=(x_{max}-x_{min})/26$, where $x_{min}$ and $x_{max}$ are the extreme values of the time series. The letters occupy the bins in a secret order, the trivial choice is the ascending order, i.e., bin(1)=A, bin(2)=B, ..., bin(26)=Z. 

Considering the ascending order, if we intend to encrypt the letter A, i.e., $m=A$, then, after $N$ iteration times the variable $x_{n'+N}$ needs to be in the interval $[x_{min},x_{min}+\epsilon)$, for $m=B$, then, $x_{n'+N} \in [x_{min}+\epsilon,x_{min}+2\epsilon)$, and so on. Where $n'$ is considered the initial time. 

To send a secret message to someone else is necessary to define a secret order for the alphabet, a secret value for the parameter $a$, and find the number of iterations $N$ such the system evolves to the desired interval. For instance, if Alice intends to send the secret message "hi" to Bob. One possibility is to adjust the model such the alphabet is in ascending order, $a=3.98798600000000$ and the initial condition is $x_0=0.01010101010101$. Skipping the 250 firsts iterations as transient time, to get the letter 'h', that is the 8th in the English alphabet, is necessary to evolve the system for 259 times, i.e., $N=259$ and $n'=0$. Again skipping the 250 firsts iterations as transient time, to get the letter 'i', that is the 9th in the English alphabet, is necessary to evolve the system more 407 times, i.e., $N=407$ and $n'=259$. Bob receives the cyphertext : $259$ $407$, if he knows the initial condition, the $a$ parameter, and the alphabet order, he can recover the message. 

\section{Our Algorithm}

\begin{figure}[htb!]
\centering
\includegraphics[width=0.7\linewidth]{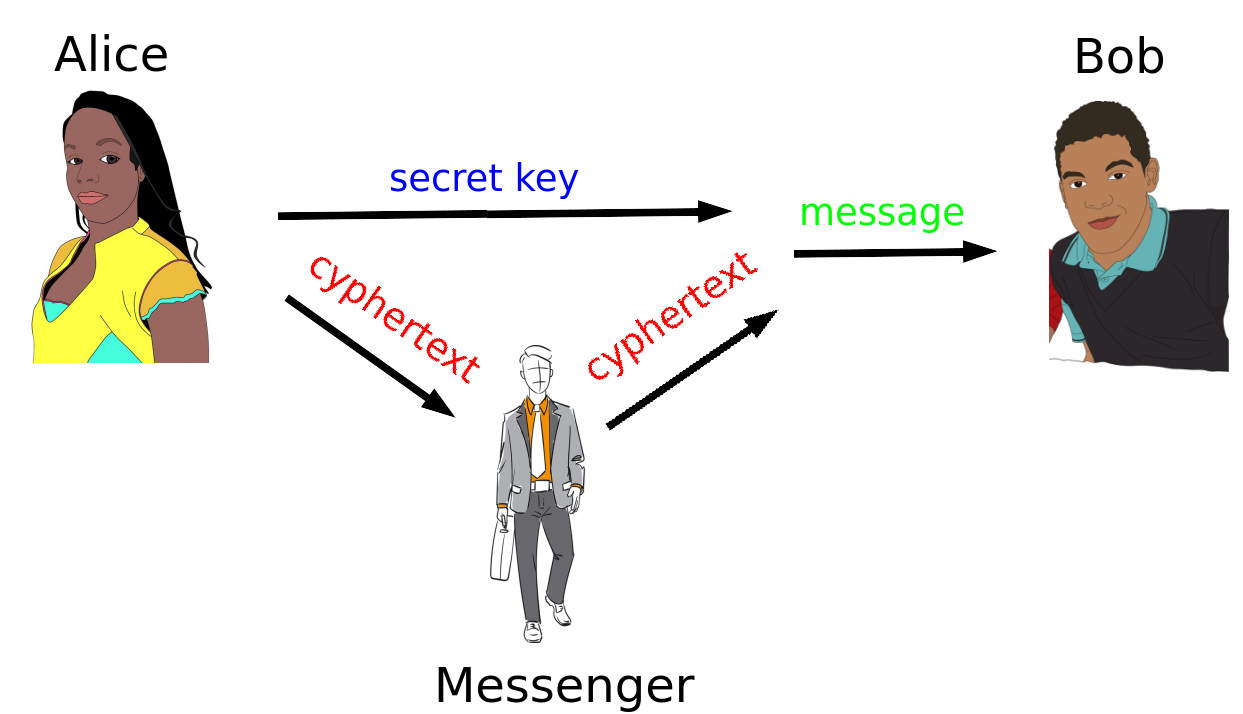}
\caption{Alice and Bob's problem.}
\label{fig1}
\end{figure}

In this section we present an alternative way to cypher messages. As illustrated in Fig. \ref{fig1}, supposing Alice is trying to send a secret message to Bob. Assuming the only way the message can be delivered is through a messenger. In order to do that, they define a secret key that is only shared between Alice and Bob. Alice encrypts the message into a cyphertext and hand it to the messenger. When Bob receives the cyphertext he uses his secret key and recovers the message. 

In order to achieve their goal they use the following protocol:

\begin{enumerate}
\item First, Alice selects a binary message: $m=0$ or $m=1$. 
\item Then, Alice defines a secret key whose size is $10^{14}$, for example, key$=12345678901234$.
\item To generate the cyphertext, Alice uses the secret key as the initial condition of a chaotic map $e_{n+1}=f(e_{n})$, in this case $f()$ is the logistic map,
\begin{eqnarray}
e_{n+1}=f(e_n )=a e_n (1-e_n ),
\end{eqnarray}
where $n$ is a discrete time step and $a$ is a fixed parameter. For this example, the initial condition will be $e_0=0.12345678901234$.
\item To encode the message in the cyphertext, Alice generates a time series, such that
\begin{eqnarray}
s_{n+1}=(1-\delta)f(s_n )+\delta m f(e_n),
\end{eqnarray}
where $s_0$ can be randomly defined. The coupling strength needs to be a value where the system is coupled but not synchronized. 
\item Alice sends $10000$ iterations of the time series through the messenger to Bob. 
\item Bob receives the time series of $s$ and generates the time series of the variable $e$ using his secret key. 
\item Bob uses a network quantifier to detect the presence or absence of coupling between the variables $s$ and $e$. 
\item When the variables are coupled the recovered message is '1', when the variables are uncoupled the recovered message is '0'. 
\end{enumerate}

Due to the sensibility of initial conditions, in this model all the time series are truncated in $10^{-14}$.

\section{Detecting coupling between variables}

The message can be recovered by means of quantities that are able to detect coupling between chaotic variables. Based on previous results, three quantities that are able to detect coupling are: Correlation Coefficient $CC$ \cite{Rubido14}, Mutual Information $MI$ \cite{Rubido14} and Transfer Entropy $TE$ \cite{Schreiber01}. 

Correlation Coefficient $CC$, in this case, Pearson's Correlation Coefficient, is a well known quantity that identifies linear correlation between variables. Considering the time series of size $N$, $x_1,x_2,...x_N$ and $y_1,y_2,...,y_N$, the Correlation Coefficient $CC$ can be defined as
\begin{eqnarray}
CC&=&\frac{\sum(x_n-\langle x \rangle)}{\sigma_x}\frac{\sum(y_n-\langle y \rangle)}{\sigma_y},
\end{eqnarray}
where $ \langle \rangle$ indicates the average, $\sigma$ is the standard deviation and the sums are over all the time series.

Mutual Information $MI$ is a measure of dependence between two variables. Considering two variables $x$ and $y$ the Mutual Information $MI$ is given by
\begin{eqnarray}
MI&=&\sum p(x_n, y_n) \log \frac{p(x_n, y_n)}{p(x_n) p(y_n)},
\end{eqnarray}
where $p()$ indicates the probabilities and the sum is over all possible states.

Transfer Entropy $TE$ is a non-symmetric quantity that allows to detect transfer of information between variables. For example, the amount of transfer entropy from variable $x$ to variable $y$ can be evaluated as
\begin{eqnarray}
TE_{X \rightarrow Y}&=& \sum {p(y_{n+1}, y_n, x_n) \log \frac{p(y_{n+1}|y_n,x_n)}{p(y_{n+1}|y_n)}},
\end{eqnarray}
where the sum is over all possible states, $p()$ and $p(|)$ are probabilities and conditional probabilities, respectively. 

\begin{figure}[htb!]
\centering
\includegraphics[width=0.7\linewidth]{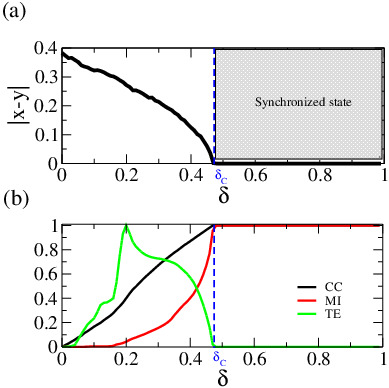}
\caption{(a) Difference between the two coupled variables and (b) normalized value of the quantifiers, as function of $\delta$. The dashed line indicates the critical value $\delta_c$ of the synchronized state. }
\label{fig2}
\end{figure}

These quantities can be tested considering a master-slave coupling between two logistic maps,
\begin{eqnarray}
\left\{\begin{array}{rcl}
x_{n+1}&=&f(x_n )\\
y_{n+1}&=&(1-\delta)f(y_n )+\delta f(x_n)
\end{array}\right..\label{eq1}
\end{eqnarray}
In this configuration, when the coupling strength $\delta$ is increased above a critical value $\delta_c$ the system evolves to a synchronized state. This state can be identified by the case where $|x-y|=0$. As shown in Fig. \ref{fig2} (a), considering $a=3.987986$, as $\delta$ is increased the difference between the variables $x$ and $y$ is reduced and goes to 0 when $\delta_c=0.47$ (blue dashed line in Fig. \ref{fig2} (a)). 

Each network quantifiers vary differently, as shown in Fig. \ref{fig2} (b). The Correlation Coefficient $CC$ (black line in Fig. \ref{fig2} (b)) increases with $\delta$ and reaches its maximum value, $CC_{\mbox{MAX}}=1$, when the variables are synchronized. The Mutual Information $MI$ (red line in Fig. \ref{fig2} (b)), increases with $\delta$ but slower than $CC$, and reaches its maximum value, $MI_{\mbox{MAX}}=\ln 2$, in the synchronized state. The Transfer Entropy $TE$ behaves in a more complicated manner, it reaches its maximum value, $TE_{\mbox{MAX}}=0.20$, when $\delta \approx 0.2$, and decreases  up to 0 in the synchronized state.

These quantifiers are able to differentiate coupled from uncoupled variables when $\delta < \delta_c$. This is possible because below $\delta_c$ the quantifiers are always a positive quantity and when $\delta=0$ the quantifiers are approximately 0. The quantities are never exactly zero because of the statistical fluctuations. The fluctuations occurs because the measures are realized far away of the thermodynamic limit (where the size of the time series is infinity). To avoid false positives or false negatives is necessary to define a threshold. 

\section{Example}

\begin{figure}[htb!]
\centering
\includegraphics[width=0.7\linewidth]{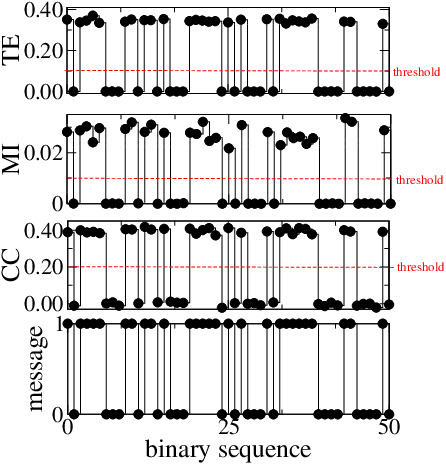}
\caption{Example of a 50 bits message. (a) $TE$ measure, (b) $MI$ measure, (c) $CC$ measure, (d) binary message. The red dashed lines indicate the threshold values.}
\label{fig3}
\end{figure}

We simulate the transmission of a 50 bits message from Alice to Bob. In Fig. \ref{fig3}, the message is randomly selected and the coupling strength is kept fixed, $\delta=0.2$. To differentiate the states '0' and '1' the standard deviation of a set of 1000 random messages is used as threshold. When a given quantity is above the threshold the result is '1' and bellow the threshold the result is '0'. We find the following threshold values: 0.1 for Transfer Entropy $TE$, 0.015 for Mutual Information $MI$ and 0.2 for Correlation Coefficient $CC$. 

\section{Robustness and Security}

\begin{figure}[htb!]
\centering
\includegraphics[width=0.7\linewidth]{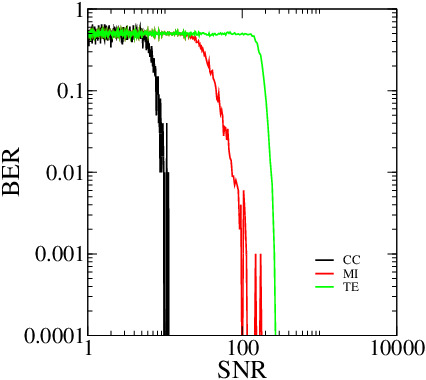}
\caption{Bit Error Rate (BER) as function of the Signal-Noise-Ratio (SNR).}
\label{fig4}
\end{figure}

In the previous example we assume that the message is delivered without errors. However, during the transmission of the message some unexpected errors can occur. To verify the robustness of the system we add a Gaussian noise in the cyphertext with null mean and variance $\sigma^2$. To quantify the robustness of the system we use the Bit-Error-Rate (BER), that is the rate of wrong bits as function of Signal-Noise-Ratio (SNR), in this case, calculated in terms of the signal amplitude $A_{\mbox{signal}}$ and noise amplitude $A_{\mbox{noise}}$,
\begin{eqnarray}
SNR=\left (\frac{A_{\mbox{signal}}}{A_{\mbox{noise}}} \right )^2.
\end{eqnarray}
The results are shown in Fig. \ref{fig4}. Our method tolerates noises up to a certain amplitude value, above it the system starts to provide bad results in an abrupt transition. Correlation Coefficient $CC$ is the quantifier that tolerates the larger values for noise amplitude and Transfer Entropy $TE$ is the less robust.                    

Despite of being robust to noise, unfortunately, our model fails against some traditional hack attacks. We are going to discuss the vulnerabilities against two types of attacks: Cyphertext Only Attack (COA) and Brutal Force Attack (BFA). 

\subsection{Cyphertext Only Attack (COA)}

\begin{figure}[htb!]
\centering
\includegraphics[width=0.7\linewidth]{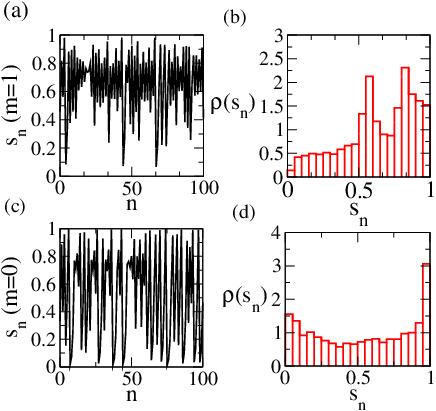}
\caption{Comparing the two possible cyphertexts. (a) Cyphertext for $m=1$, (b) probability distribution for $m=1$, (c) cyphertext for $m=0$, (d) probability distribution for $m=0$.}
\label{fig5}
\end{figure}

In this type of attack the hacker has only access to the cyphertext. Our method works with binary states, in Fig. \ref{fig5} we analyze the probability distribution of the time series considering the two possible messages: '1' or '0'. It is not possible to extract information when we compare Fig. \ref{fig5} (a) and (c), however when we look to the probability distribution of the cyphertexts, Fig. \ref{fig5} (b) and (d), they are very distinguishable. 

\subsection{Brutal Force Attack (BFA)}

The Brutal Force Attack (BFA) consists of trying all the possible keys to decrypt the message. The vulnerability in this cases consists of the truncation of the time series. When the time series is truncated in $10^{-14}$ there are $10^{14}$ possible keys. The number of possible keys is huge but is easily achievable with a personal computer. 

\section{Final Considerations}

We presented an algorithm to cypher messages using the coupling between chaotic maps. We show that the method is stable and robust against noise. Unfortunately, after several analyzes our method fails to be considered secure against classical hacker attacks. 

On the other hand, after twenty years, the Baptista's algorithm presented in Section 2.3 still remains as reference for chaos based cryptography. In 2014, Li {\it et al.} presented modifications in the model to improve it against several types of attacks \cite{Li2004}.

Despite our lack of success, all cryptography methods are either vulnerable or computationally inefficient in some manner. Despite of the difficulties, the continuous advance of computation, cryptocurrencies and artificial intelligence demand new methods of protection. 

We thanks CNPq for partial financial support.

\label{pagefin}
\end{document}